\documentclass[useAMS,usenatbib]{mn2e} \usepackage{amsmath,amssymb}
\usepackage{graphicx,mathptm} \usepackage{times} \usepackage{natbib}
\usepackage{psfig}

\newcommand{\msun}{${\rm M}_{\odot}$\,}
\newcommand{\kpch}{$h^{-1}\,\mbox{kpc}$\,}
\newcommand{\Mpch}{$h^{-1}\,\mbox{Mpc}$\,}
\newcommand{\ergcms}{$\mbox{erg}\,\mbox{cm}^{-2}\,\mbox{s}^{-1}$}
\title[AGN in mock Chandra deep fields]
{The spatial distribution of X-ray selected AGN in the {\em Chandra}
  deep fields: a theoretical perspective}
\author[Marulli et al.]
{
Federico Marulli$^1$, 
Silvia Bonoli$^2$,
Enzo Branchini$^3$,
Roberto Gilli$^4$,
Lauro Moscardini$^{1,5}$ 
\newauthor
and Volker Springel$^2$
\\ 
$^1$Dipartimento di Astronomia, Universit\`a degli Studi di Bologna, 
via Ranzani 1, I-40127 Bologna, Italy \\
$^2$Max-Planck-Institut f\"ur Astrophysik, Karl-Schwarzschild Strasse 1,
D-85740 Garching, Germany\\
$^3$Dipartimento di Fisica, Universit\`a 
degli Studi ``Roma Tre'', via della Vasca Navale 84, I-00146 Roma,
Italy\\
$^4$Istituto Nazionale di Astrofisica (INAF) - Osservatorio
Astronomico di Bologna, Via Ranzani 1, 40127 Bologna, Italy\\ 
$^5$INFN, Sezione di Bologna, viale Berti Pichat 6/2, I-40127 Bologna, Italy
}

\begin{document}

\maketitle

\label{firstpage}


\begin{abstract}
  We study the spatial distribution of X-ray selected AGN in the
  framework of hierarchical co-evolution of supermassive black holes
  (BHs) and their host galaxies and dark matter (DM) haloes. To this
  end, we have applied the theoretical model developed by
  \citet{croton2006}, \citet{delucia2007} and \citet{marulli2008} to
  the output of the Millennium Run and obtained hundreds of
  realizations of past light-cones from which we have extracted
  realistic mock AGN catalogues that mimic the {\em Chandra} deep
  fields.  We find that the model AGN number counts are in fair
  agreement with observations both in the soft and in the hard X-ray
  bands, except at fluxes $\lesssim10^{-15}$ \ergcms, where the model
  systematically overestimates the observations.  However, a large
  fraction of these faint objects is typically excluded from the
  spectroscopic AGN samples of the {\em Chandra} fields. We find that
  the spatial two-point correlation function predicted by the model is
  well described by a power-law relation out to 20 \Mpch, in close
  agreement with observations. Our model matches the correlation
  length $r_0$ of AGN in the {\em Chandra} Deep Field North but
  underestimates it in the {\em Chandra} Deep Field South.  When
    fixing the slope to $\gamma = 1.4$, as in \citet{gilli2005}, the
    statistical significance of the mismatch is 2-2.5 $\sigma$,
    suggesting that the predicted cosmic variance, which dominates the
    error budget, may not account for the different correlation length
    of the AGN in the two fields.  However, the overall mismatch
    between the model and the observed correlation function decreases
    when both $r_0$ and $\gamma$ are allowed to vary, suggesting that
    more realistic AGN models and a full account of all observational
    errors may significantly reduce the tension between AGN clustering
    in the two fields.  While our results are robust to changes in
  the model prescriptions for the AGN lightcurves, the luminosity
  dependence of the clustering is sensitive to the different
  lightcurve models adopted. However, irrespective of the model
  considered, the luminosity dependence of the AGN clustering in our
  mock fields seems to be weaker than in the real {\em Chandra}
  fields. The significance of this mismatch needs to be confirmed
  using larger datasets.
\end{abstract}


\section {Introduction}

A cosmological co-evolution of DM structures, galaxies and BHs is
expected within the standard $\Lambda$CDM framework \citep[see,
  e.g.][and references therein]{volonteri2003a, cattaneo2005,
  marulli2006, croton2006, fontanot2006, malbon2007, hopkins2008,
  marulli2008} and strongly supported by several observational
evidences like, for example, the BH scaling relations and the
luminosity function of galaxies and AGN \citep[see,
  e.g.][]{magorrian1998,tremaine2002,marconi2003,ferrarese2005,hopkins2007,graham2008}.
Modelling these observations is a significant challenge for modern
computational astrophysics, as it requires to self-consistently
account for complex physical processes acting both on very large
scales, like the ones related to galaxy formation and evolution, and
on very small scales, like the gas cooling and the mass accretion onto
the central BHs.

The computational cost of full cosmological hydrodynamical simulations
is very high, and only few attempts have been made thus far to
directly follow the co-evolution of BHs and their host galaxies within
large cosmological volumes from high redshifts to the present epoch
\citep{li2007,pelupessy2007,sijacki2007,dimatteo2008}. Moreover, every
modification of the prescriptions used to encapsulate the `sub-grid'
physics requires the simulations to be repeated.  A popular,
considerably less time consuming alternative is to run
high-resolution, cosmological simulations of the DM component alone
and apply semi-analytic prescriptions in post-processing to model the
diffuse galactic gas and its accretion onto the central BH.  Using
this `hybrid' approach, a galaxy formation model has been implemented
on top of the Millennium Run \citep{springel2005}, a very large
simulation of the concordance $\Lambda$CDM cosmology, which follows
the DM evolution from $z=127$ to the present, in a comoving box of 500
\Mpch on a side and with a comoving scale resolution of 5 \kpch. The
galaxy formation model has been originally proposed by
\citet{springel2001b} and \citet{delucia2004b} and subsequently
updated to include a `radio mode' BH feedback \citep{croton2006,
  delucia2007} and to self-consistently describe the BH mass accretion
rate triggered by galaxy merger events (`quasar' mode) and its
conversion into radiation \citep[][hereafter M08]{marulli2008}.  The
model outputs are publicly available at the Millennium download site
at the German Astrophysical Virtual
Observatory\footnote{http://www.g-vo.org/Millennium}
\citep{lemson2006}.

Here, we use an updated version of the model as presented in M08.  In
several previous works the model has been extensively compared to a
large set of observational data. Thanks to the `radio mode' BH
feedback, the model is able to reproduce the observed low mass
drop-out rate in cooling flows, the exponential cut-off in the bright
end of the galaxy luminosity function and the bulge-dominated
morphologies and old stellar ages of the most massive galaxies in
clusters \citep{croton2006}. In fact, model predictions are in
  agreement with several different properties of the galaxy and BH
  populations \citep[see e.g.][and reference therein]{delucia2004,
    delucia2006, springel2005, wang2007, croton2008, delucia2008}.
In M08 the model predictions have been compared to the observed
scaling relations, fundamental plane and mass function of BHs, and to
the luminosity function of AGN. The agreement between predicted and
observed BH properties is generally quite good. Also, the AGN
luminosity function can be well matched over the whole redshift range,
provided it is assumed that the cold gas fraction accreted by BHs at
high redshifts is larger than at low redshifts. Despite this
  success, some authors found discrepancies between model predictions
  and some observations \citep[see e.g.][]{weinmann2006,
    kitzbichler2007, elbaz2007, mccracken2007, gilli2007b,
    mateus2008}.  This suggests that several improvements in the
physical assumptions of the semi-analytic model are needed to make the
model predictions agree closer with these observations.  However, this
is beyond the scope of the present work, in which we focus on studying
the present model predictions about the BH and AGN populations,
extending the analysis of M08.

In this work, we focus on the AGN clustering, which represents an
additional, fundamental observational property that provides further
constraints to the theoretical models. Together with the AGN
luminosity function, the galaxy mass function and their bias, the AGN
clustering can be used to constrain the masses of the AGN host
galaxies, and thus the AGN lifetimes. In fact, if AGN are long-lived
sources, then they are probably rare phenomena occurring in massive
haloes, highly biased with respect to the underlying mass
distribution. On the contrary, if they are short-lived they likely
reside in typical haloes that are less clustered than the massive
ones.

 In recent years, wide-field surveys of optically selected AGN have
 enabled tight measurements of the unobscured (type-1) AGN clustering
 up to $z\sim3$ \citep[see e.g.][]{porciani2004, grazian2004,
   croom2005, porciani_norberg2006}.  The use of X-ray selected AGN
 catalogues allows one to include also obscured (type-2) objects, thus
 minimizing the impact of bolometric corrections. However, such
 observational studies have been limited by the lack of sizeable
 samples of optically identified X-ray sources.  To overcome this
 problem, \citet{gilli2005} used the two deepest X-ray fields to date,
 i.e. the 2Msec {\em Chandra} Deep Field North
 \citep[CDFN,][]{alexander2003,barger2003a} and the 1Msec {\em
   Chandra} Deep Field South
 \citep[CDFS,][]{rosati2002,giacconi2002}\footnote{The CDFS exposure
   has been recently extended to 2 Msec, and an updated X-ray
   catalogue has been already released \citep{luo2008}. In this work,
   however, we will keep working with the 1Msec X-ray source catalogue
   of \citet{giacconi2002}, for which optical identification is almost
   complete.}.  Limiting fluxes (in \ergcms) of
 $\sim2.5\times10^{-17}$ and $\sim1.4\times10^{-16}$ for the CDFN and
 of $\sim5.5\times10^{-17}$ and $\sim4.5\times10^{-16}$ for the CDFS
 have been reached in the soft (0.5-2 keV) and hard (2-10 keV) X-ray
 bands, respectively.  A sample of 503 sources in the CDFN and 346
 sources in the CDFS has been collected over two areas of 0.13 and 0.1
 ${\rm deg}^2$, respectively.  The correlation properties of the AGN
 in these two fields turned out to be quite different since the
 correlation length, $r_0$, measured in the CDFS is a factor of
 $\sim2$ higher than in the CDFN \citep{gilli2005}. As it seems
 unlikely that this difference can be due only to observational
 biases, it has been argued that it could be accounted for if one
 includes the cosmic variance, supposedly large in these deep fields,
 in the error budget.

To successfully discriminate between different AGN models one needs to
account for all possible systematic errors that may plague the
comparison between theoretical predictions and observations.  For this
purpose, we construct a large set of mock AGN catalogues that mimic as
close as possible the observed properties of the X-ray selected AGN in
the two {\em Chandra} fields and account for all known observational
biases.  We then use these simulated samples to `observe' the number
counts of mock AGN and their clustering properties that we then
compare to observations.  Thanks to the large box of the Millennium
Simulation where many such independent samples can be extracted from,
we can directly assess the impact of the cosmic variance by measuring
the field-to-field variation of the mock AGN clustering properties.

The paper is organized as follows. In Section \ref{sec:model}, we
briefly discuss the main aspects of the hybrid simulation used to
construct the mock AGN catalogues.  In Section \ref{sec:fields}, we
describe the technique used to extract realistic mock {\em Chandra}
fields from the Millennium Simulation.  We compare the predicted AGN
number counts and spatial clustering with those measured in the {\em
  Chandra} Deep Fields in Section \ref{sec:comparison}. Finally, in
Section \ref{sec:conclusions}, we summarize our conclusions and
discuss our results.


\section {The AGN model} \label{sec:model}

The hybrid simulation used in this paper is described in detail in
\citet{croton2006} and \citet{delucia2007}. In the following, we just
give a brief description of the main features of the model and review
the new semi-analytic recipes recently included by M08 to describe the
AGN evolution.

\subsection {DM haloes and galaxies}

The model simulates the co-evolution of DM haloes, galaxies and their
central BHs in the $\Lambda$CDM `concordance' cosmological framework,
with parameters $\Omega_{\rm m}=0.25$, $\Omega_{\rm b}=0.045$,
$\Omega_\Lambda=0.75$, $h=H_0/100\, {\rm km\, s^{-1} Mpc^{-1}}=0.73$,
$n=1$, and $\sigma_8=0.9$, consistent with determinations from the
combined analysis of the 2-degree Field Galaxy Redshift Survey
(2dFGRS) \citep{colless2001} and first-year WMAP data
\citep{spergel2003}, as shown by \cite{Sanchez2006}. The DM evolution
is described through a numerical N-body simulation, the Millennium
Run, which followed the dynamics of $2160^3\simeq 10^{10}$ DM
particles with mass $8.6\times10^8\,h^{-1}{\rm M}_{\odot}$ in a
periodic box of $500\,h^{-1}$Mpc on a side \citep{springel2005}.

The baryonic physics is implemented in a post-processing phase, 
  by exploiting the merging trees of DM haloes extracted from the
  simulation. Two different techniques have been used to identify DM
  haloes and their substructures: the friends-of-friends (FOF)
  group-finder and an updated version of the {\small SUBFIND}
  algorithm \citep{springel2001b}. To establish the baryons to DM halo
  connection we assume that, when DM haloes collapse a fixed mass
  baryon fraction collapses along, as proposed by \citet{white1991}.
  The baryon component, initially in the form of diffuse, pristine
  gas, forms stars and change its chemical composition.  The evolution
  of this diffuse gas is regulated by heating and cooling processes
  described by using physically motivated prescriptions.  The
photo-ionization heating of the intergalactic medium is invoked to
suppress the concentration of baryons in shallow potentials
\citep{efstathiou1992} and to make the accretion and cooling in
low-mass haloes inefficient.  The star formation rate is assumed
  to be proportional to the cold gas mass of the galaxy, while the
  supernovae reheating of the hot interstellar gas medium is
  proportional to the mass of stars. If an excess of SN energy
  is present after reheating material to the halo virial temperature,
  then an appropriate amount of gas leaves the DM halo in the form of
  a `super-wind'.  Galaxy disk instability is modelled using the
analytic stability criterion of \citet{mo1998}.  DM substructures are
followed until tidal truncation and stripping disrupt them, or they
fall below a mass of $1.7\times 10^{10}h^{-1}M_\odot$.  At this point,
a survival time is estimated using the subhalo's current orbit and the
dynamical friction formula of \citet{binney1987} multiplied by a
factor of $2$, as in \citet{delucia2007}. After this time, the galaxy
is assumed to merge onto the central galaxy of its own halo.  The
  starburst triggered by galaxy mergers is modelled with the
  prescriptions introduced by \citet{somerville2001}.

\begin{figure}
\includegraphics[width=0.48\textwidth]{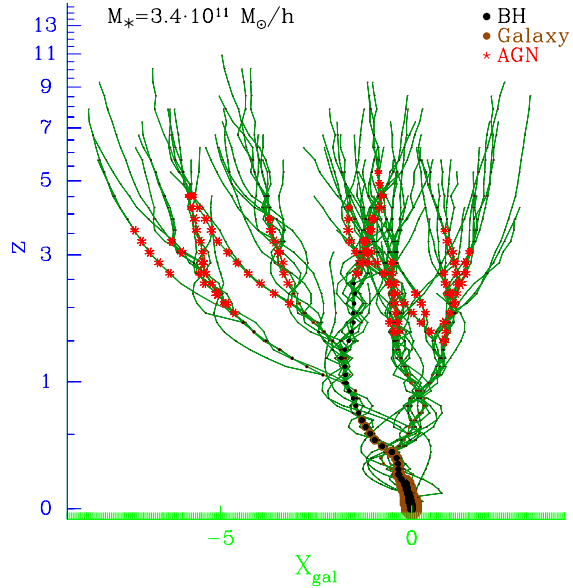}
\caption{ A typical galaxy merger tree in our model. The sizes of
  brown and black dots are proportional to the stellar mass of the
  galaxies and to the mass of the central BHs, respectively. The red
  stars indicate the presence of an AGN and their sizes are
  proportional to the AGN bolometric luminosities.  In the example
  shown, the merging history of a parent galaxy with stellar mass
  $M_\star=3.4\cdot10^{11}\,h^{-1}$ \msun is traced back in time from
  $z=0$, at the bottom of the plot, out to $z \sim 10$.  The
    variable on the horizontal axis represents the displacement
    between the parent galaxy and its progenitor, defined as $X_{\rm
      gal}=\sum^3_{i=1} (x^i_{\rm gal}-x^i_{\rm par})$, where
    $x^i_{\rm gal}$ and $x^i_{\rm par}$ represent the three Cartesian,
    comoving components of the progenitor and the parent galaxy,
    respectively, in unit of \Mpch.}  
 \label{fig:merger}
\end{figure}

In Fig.~\ref{fig:merger} we show a typical merger tree in our model.
The sizes of brown and black dots are proportional to the stellar mass
of the galaxies and to the mass of the central BHs, respectively. The
red stars indicate the presence of an AGN and their sizes are
proportional to the AGN bolometric luminosities.  In the example
shown, the merging history of a parent galaxy with stellar mass
$M_\star=3.4\cdot10^{11}\,h^{-1}$ \msun is traced back in time from
$z=0$, at the bottom of the plot, out to $z \sim 10$.

\subsection {Supermassive black holes}\label{subsec:BH}

In order to populate our model galaxies with BHs and AGN, we adopt the
following assumptions.  The BH mass accretion is triggered by two
different phenomena: i) the merger between gas-rich galaxies and ii)
the cooling flow at the centres of X-ray emitting atmospheres in
galaxy groups and clusters.

The first kind of accretion, dubbed {\em quasar mode}, is closely
associated with starbursts. Many recent works seem to indicate
  that major mergers do not constitute the only trigger to BH
  accretion \citep[see e.g.][and reference therein]{marulli2007,
    kauffmann2008, hopkins2008b, silverman2008}.  For this reason, we
  assume here that {\it any} galaxy merger can trigger perturbations
  to the gas disk and drives gas onto the galaxy centre. BHs can
  accrete mass both through coalescence with another BH and by
  accreting cold gas, the latter being the dominant accretion
  mechanism.  The gas mass accreted during a merger is assumed to be
proportional to the total cold gas mass of the galaxy
\citep{Kauffmann2000}, but with an efficiency which is lower for
smaller mass systems and for unequal mergers:
\begin{equation}
\Delta M_{\rm BH} = f_{\rm BH} \frac{m_{\rm sat}}{m_{\rm central}}
\frac{m_{\rm cold}}{1 + V_{\rm vir, 280}^{-2}}\, ,
\label{eq:accretionQ}
\end{equation}
where $m_{\rm sat}/m_{\rm central}$ is the total mass ratio of merging
galaxies, $m_{\rm cold}$ and $V_{\rm vir, 280}$ are the cold gas mass
and the virial velocity (in units of $280\,{\rm km\,s^{-1}}$) of the
central galaxy, respectively. The parameter $f_{\rm BH}\approx 0.03$
is chosen to reproduce the observed local $M_{\rm BH}-M_{\rm bulge}$
relation \citep{croton2006}. The accretion driven by major mergers is
the dominant mode of BH growth in this scenario.  Its energy feedback,
which has not been included in the model so far, is approximated by an
enhanced effective feedback efficiency for the supernovae associated
with the starburst.

Once a static hot halo is formed around a galaxy, we assume that the
{\em radio mode} sets in, in which a fraction of the hot gas
quiescently accretes onto the central BH.  During this phase, the
accretion rate is typically orders-of-magnitude below the Eddington
limit, so that the growth of the BH mass is negligible compared to
during the {\em quasar mode} phase.  However, the energy feedback
associated with it injects enough energy into the surrounding medium
to reduce or even stop the cooling flow in the halo centres.  In this
scenario, the effectiveness of radio AGN in suppressing cooling flows
is greatest at late times and for large values of the BH mass.

The mass accretion onto the BHs and the associated bolometric
luminosity emitted can be described as follows:
\begin{equation} 
  L_{\rm bol}(t) = f_{\rm Edd}(t)L_{\rm Edd}(t) 
\end{equation} 
\begin{equation} 
  \frac{d\ln M_{\rm BH}(t)}{dt} = {t_{\rm ef}^{-1}(t)}
\end{equation} 
where $L_{\rm Edd}$ is the Eddington luminosity, $t_{\rm
  ef}(t)=\frac{\epsilon}{1-\epsilon}\frac{t_{\rm Edd}}{f_{\rm
    Edd}(t)}$ is the e-folding time ($t_{\rm ef}\equiv t_{\rm
  Salpeter}$ if $f_{\rm Edd}=1$), $\epsilon$ is the {\em radiative
  efficiency}, $f_{\rm Edd}(t)$ is the {\em Eddington factor} and
$t_{\rm Edd}=\sigma_{T} c /(4\pi m_{p} G) \sim0.45\,{\rm Gyr}$.  As in
M08, we do not follow the evolution of the BH spins and we take a
constant mean value for the radiative efficiency of $\epsilon=0.1$ at
all redshifts.

We consider three different prescriptions to model $f_{\rm Edd}$,
which determines the lightcurves associated with individual quasar
events:

\begin {itemize}

\item {\em I}: $f_{\rm Edd}=1$, the simplest possible assumption.

\item {\em II}: $f_{\rm Edd}$ is assumed to decrease at low $z$ as
  suggested by \citet{cattaneo2003} and \citet{shankar2004} to match
  the BH mass function derived from a deconvolution of the AGN
  luminosity function and the local BH mass function.  Here, we adopt
  the fit derived by \citet{shankar2004}:
  \begin{equation}
    f_{\rm Edd}(z)=\left\{
    \begin{array}{ll}
      0.3       &  z\geqslant3   \\
      0.3\cdot[(1+z)/4]^{1.4}      &   z<3
    \end{array}
    \right.
  \end{equation}

\item {\em III}: the evolution of an active BH is described as a
  two-stage process of a rapid, Eddington-limited growth up to a peak
  BH mass, preceded and followed by a much longer quiescent phase with
  lower Eddington ratios.  In this latter phase, the average time
  spent by AGN per logarithmic luminosity interval can be approximated
  as \citep{hopkins2005}
    \begin{equation} 
    \frac{{\rm d}t}{{\rm d}\ln L_{\rm
        bol}}=|\alpha|\,t_9\left(\frac{L_{\rm
        bol}(t)}{10^9L_\odot}\right)^\alpha,
    \label{eq:dt_hopkins} 
  \end{equation} 
  where $t_9$ is the total AGN lifetime above $10^9 L_\odot$;
  $t_9\sim10^9\,{\rm yr}$ over the range $10^9L_\odot<L_{\rm
    bol}<L_{\rm peak}$, where $L_{\rm peak}$ is the AGN luminosity at
  the peak of its activity.  In the range $10^{10}L_\odot\lesssim
  L_{\rm peak}\lesssim 10^{14}L_\odot$, \citet{hopkins2005} found that
  $\alpha$ is a function of only $L_{\rm peak}$, given by
  $\alpha=-0.95+0.32\log(L_{\rm peak}/10^{12}L_\odot)$, with
  $\alpha=-0.2$ (the approximate slope of the Eddington-limited case)
  as an upper limit. Here we interpret the Hopkins model as describing
  primarily the decline phase of the AGN activity, after the BH has
  grown at the Eddington rate to a peak mass $M_{\rm BH,peak}=M_{\rm
    BH}(t_{\rm in})+ \mathcal{F} \cdot\Delta M_{\rm
    BH,Q}\cdot(1-\epsilon)$, where $M_{\rm BH}(t_{\rm in})$ is the
  initial BH mass and $\Delta M_{\rm BH,Q}$ is the fraction of cold
  gas mass accreted.  We found that $\mathcal{F}=0.7$ is the value
  that best matches the AGN luminosity function (M08).
 
  From equation (\ref{eq:dt_hopkins}) we can derive:
\begin{equation} 
  M_{\rm BH}(t)=M_{\rm BH,peak}+\frac{A}{BC}\left[\left(1+Ct\right)^B-1\right],
  \label{eq:Mbh}
\end{equation}
where $A = \frac{1-\epsilon}{\epsilon}\frac{M_{\rm BH,peak}}{t_{\rm
Edd}}$, $B = \frac{1}{\alpha}+1$, $C = \left(\frac{L_{\rm peak}}{10^9
L_\odot}\right)^{-\alpha}\frac{1}{t_9}$.  

\end {itemize}

As shown in M08, the semi-analytic models described above
underestimate the number density of luminous AGN at high redshifts,
independently of the lightcurve model adopted. A significant
improvement can be obtained by simply assuming an accretion efficiency
that increases with redshift \citep{croton2006}. In a parallel work,
\citet{bonoli2008}, we discuss a model in which the
accretion efficiency is  linearly dependent on redshift.  In
the present work however, since our aim is to construct mock catalogues that
best reproduce the observed AGN population, we will use the model for
the accretion efficiency introduced in M08 to obtain a good match to
the AGN luminosity function:
\begin{equation}
  \left\{
    \begin{array}{ll}
      f_{\rm BH}=0.01\cdot \log\left(\frac{M_{\rm BH}}{10^3 M_\odot}+1\right)\cdot z  
      &  z>1.5 \,{\rm and}\, M_{\rm BH}>10^6M_\odot \\
      {\Delta}M_{\rm BH} = 0.01\cdot m_{\rm cold}  &  z>6
    \end{array}
    \right.
    \label{eq:best}
\end{equation}
Here we keep the prescription {\em III} for the quasar lightcurves
and, for simplicity, we assume $M_{\rm BH,seed}=10^3 M_\odot$ for all
seed BHs, irrespective of their halo host properties and their origin.
As in M08, we will refer to this scenario as our {\em best} model.
Note, however, that future improvements in the underlying physical
assumptions may well lead to a yet better model in explaining the
observations.


\section{Simulating the {\it Chandra} deep fields} \label{sec:fields}

\begin{figure*}
\includegraphics[width=\textwidth]{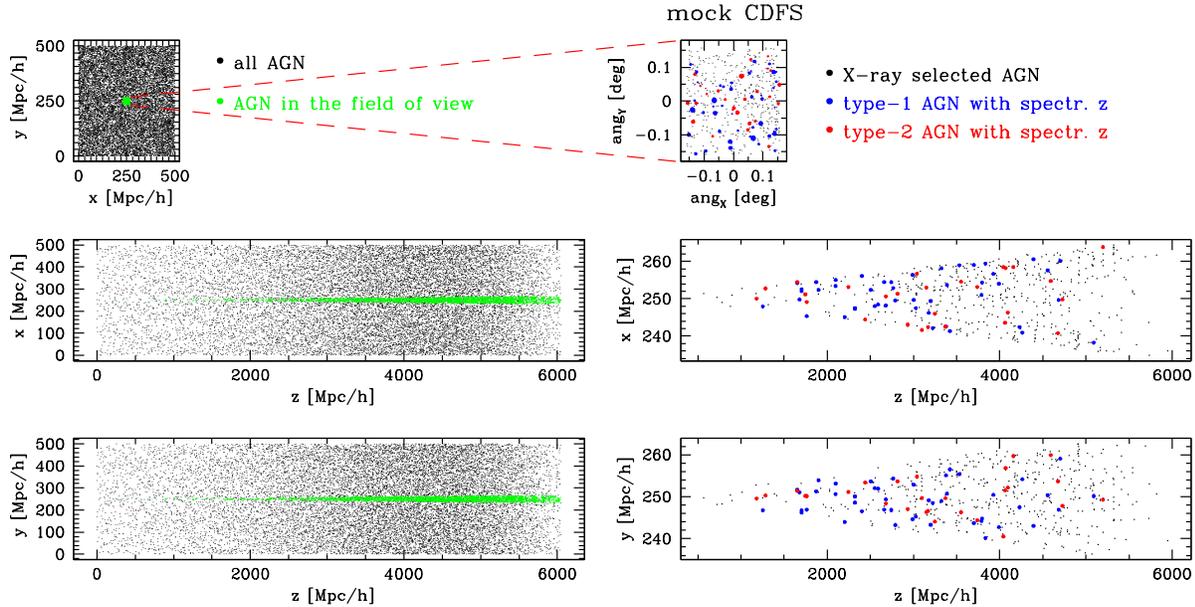}
\caption{Space-like projections of a past light cone and of a
    mock field of view with the same selection effects as the CDFS.
  Left-hand panel: all the AGN predicted by the lightcurve model I
  (black dots) inside a virtual past light cone and the subpopulation
  of AGN in a mock CDFS (green dots). Right-hand panel: zoom of the
  mock field of view represented by the green points in the left
  panel. The black dots show the AGN that meet the flux-selection
    criteria. Their number counts and redshift distributions are
    displayed in Figs.~\ref{fig:logNlogS} and \ref{fig:Zdistr},
    respectively. The blue and red dots show the type-1 and type-2 AGN
    in the mock spectroscopic subsample specified in Section
    \ref{subsec:clustering} used to compute the two-point
    correlation. The size of the red and blue dots in the upper right
    panel scales as the logarithm of the AGN observed flux.}
\label{fig:cone}
\end{figure*}

In order to directly compare our model predictions to the observed
number counts and spatial clustering of the X-ray selected AGN in the
CDFN and CDFS, we construct a suite of realistic mock AGN catalogues
that mimic the selection effects of the real data.  The aim is to
account for all uncertainties stemming from the conversions between
observed and intrinsic AGN properties and to estimate statistical
errors. Systematic errors are accounted for by modeling the AGN
  samples selection effects. Random errors contributed by sparse
  sampling in the flux limit catalogues and cosmic variance are also
  taken into account by considering several independent mock samples
  of AGN with number density comparable to that of the real {\it
    Chandra} fields.  Our realistic mock catalogs are obtained by
constructing backward light cones from the outputs of the Millennium
Simulation \footnote{A light cone is a three-dimensional hypersurface,
  in space-time coordinates, satisfying the condition that light
  emitted from every point is received by an observer at $z=0$. Its
  space-like projection is the volume of the sphere defining the
  observer's current particle horizon. The observer's field of view is
  the projection on the celestial sphere of a three-dimensional
  submanifold, in space coordinates, located inside the observer's
  particle horizon.}.  To do this, we have to take into account
that redshift varies continuously, whereas the outputs of a simulation
have been stored at a finite set of redshifts.  To interpolate between
discrete redshifts, we have used a technique similar to the standard
approach described in the literature \citep[see e.g.][]{croft2001,
  blaizot2005, roncarelli2006, kitzbichler2007}, in which the stacking
of several computational boxes corresponding to different redshift
outputs is performed in comoving coordinates.

To construct mock {\it Chandra} fields, we have considered the spatial
position and bolometric luminosity of the model AGN in the Millennium
Simulation, specified at the available output redshifts, spaced in
expansion factor according to $\log(1+z_n)=n(n+35)/4200$
\citep{springel2005}.  As a first step, we randomly locate a virtual
observer in the box at $z=0$ and transform the coordinates to have it
at the centre. Then we construct its backward light cone, which
extends to $z=5.72$, corresponding to a comoving distance of $\sim
6000$ \Mpch in our cosmological model, so one would need to stack the
simulation volume roughly 12 times. However, we can take advantage of
the much denser redshift sampling of the output times (there are $\sim
45$ different outputs between $z=0$ and $z=5.72$) by adopting the
following procedure.  We divide the light cone into slices along the
line of sight based on the output times, so that each slice
corresponds to one output and covers the redshift range closest to
this output time.  To avoid having replicas of the same cosmic
structures along the line of sight, we exploit the periodic boundary
conditions and adopt the same scrambling technique used by
\citet{roncarelli2006}. All CDFs were extracted from different
  light cones. The procedure is repeated $100$ times, for each of the
  4 lightcurve models considered and for the CDFS and CDFN separately
  (totaling to 400+400 mock CDFs samples). To perform the analysis
described in Section 4.2, it is important to estimate how many of
theses samples are statistically independent. This can be done by
comparing the volume of each sample to that of the Millennium
Simulation box, taking into account that the very rare AGN with $z>2$
do not affect the clustering property of the sample and can be safely
excluded from the spatial correlation analysis, as we did check. It
turns out that, for each lightcurve model, all the 100+100 CDFs
  extracted from the Millennium box are independent and will be
  treated as such in the rest of this work. Mock {\em Chandra} fields
are obtained by mimicking the selection effects of the real samples.
To do this, we identify all AGN with the BHs in the {\em quasar phase}
and discard those too faint to meet the flux-selection criteria.  The
latter are based on the flux measured in the soft and hard X-ray
bands, while our models predict bolometric luminosity.  To convert
intrinsic bolometric luminosities into soft and hard X-ray bands, we
use the bolometric correction proposed by \citet{hopkins2006}, which
assumes that the average AGN X-ray spectrum beyond $0.5$ keV can be
approximated by a power-law with an intrinsic photon index
$\Gamma=1.8$. To transform the intrinsic flux into the observed one,
we need to account for photon absorption along the line of sight.  To
do that, we impose that the intrinsic hydrogen column densities,
$N_{\rm H}$, of our AGN are distributed according to
\citet{lafranca2005}, and that the Galactic $N_{\rm H}$ towards the
CDFN and CDFS is $(1.3\pm0.4)\times10^{20}\mbox{cm}^{-2}$ and
$(8.8\pm0.4)\times10^{19}\,\mbox{cm}^{-2}$, respectively.  We have
checked that using the $N_{\rm H}$ distribution as proposed by
\citet{gilli2007} has a negligible effect on the final results.  Only
AGN with observed fluxes above the limit $F_{\rm limit}$ of the CDFN
and CDFS are included in our mock catalogues. The value of $F_{\rm
  limit}$ in the CDFN and CDFS varies across the field of view. We
account for this effect by adopting the dependency of $F_{\rm limit}$
from the angular distance from the fields' centre given by
\citet{giacconi2002} and \citet{bauer2004}.

\begin{figure*}
\includegraphics[width=0.48\textwidth]{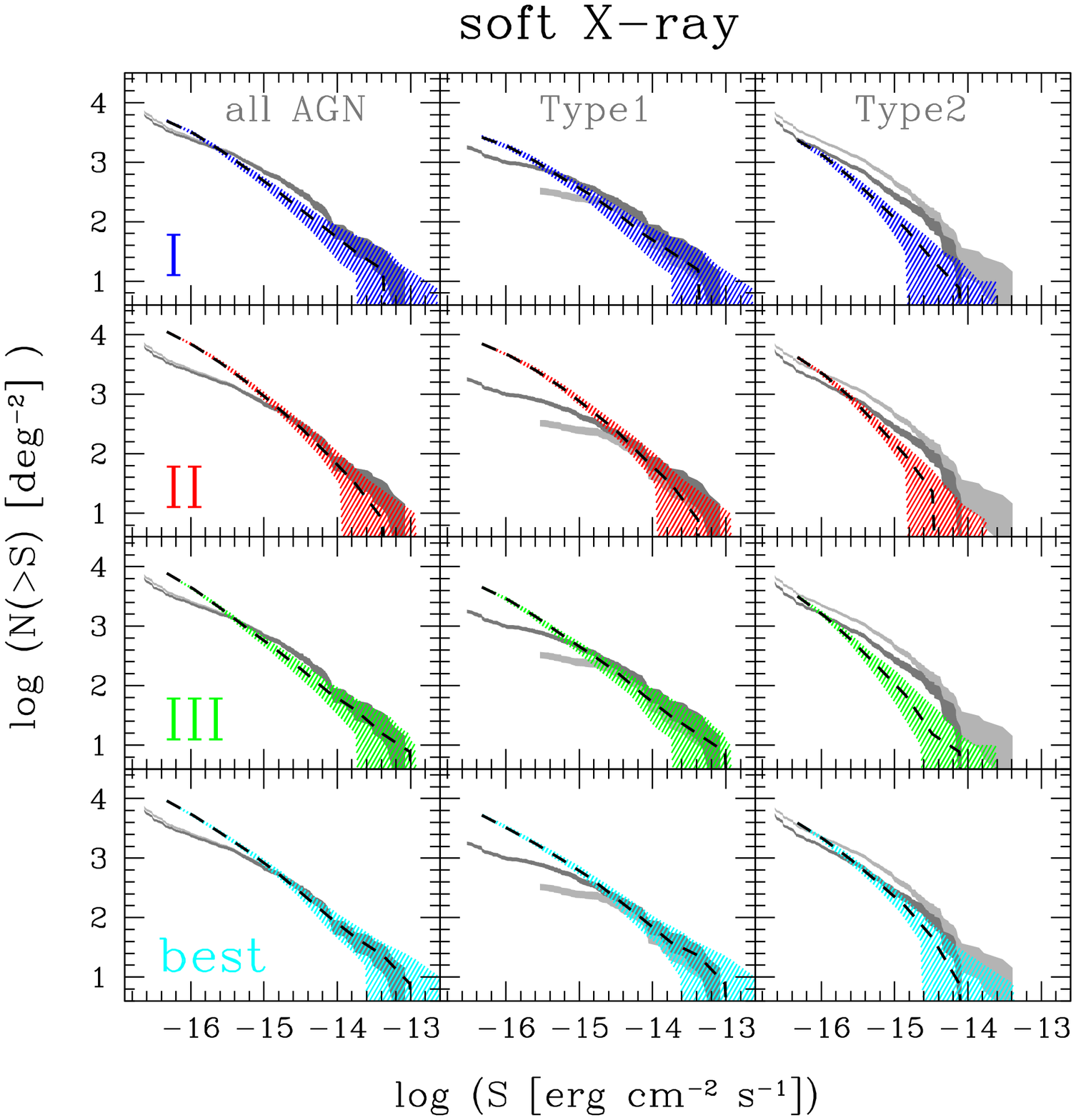}
\includegraphics[width=0.48\textwidth]{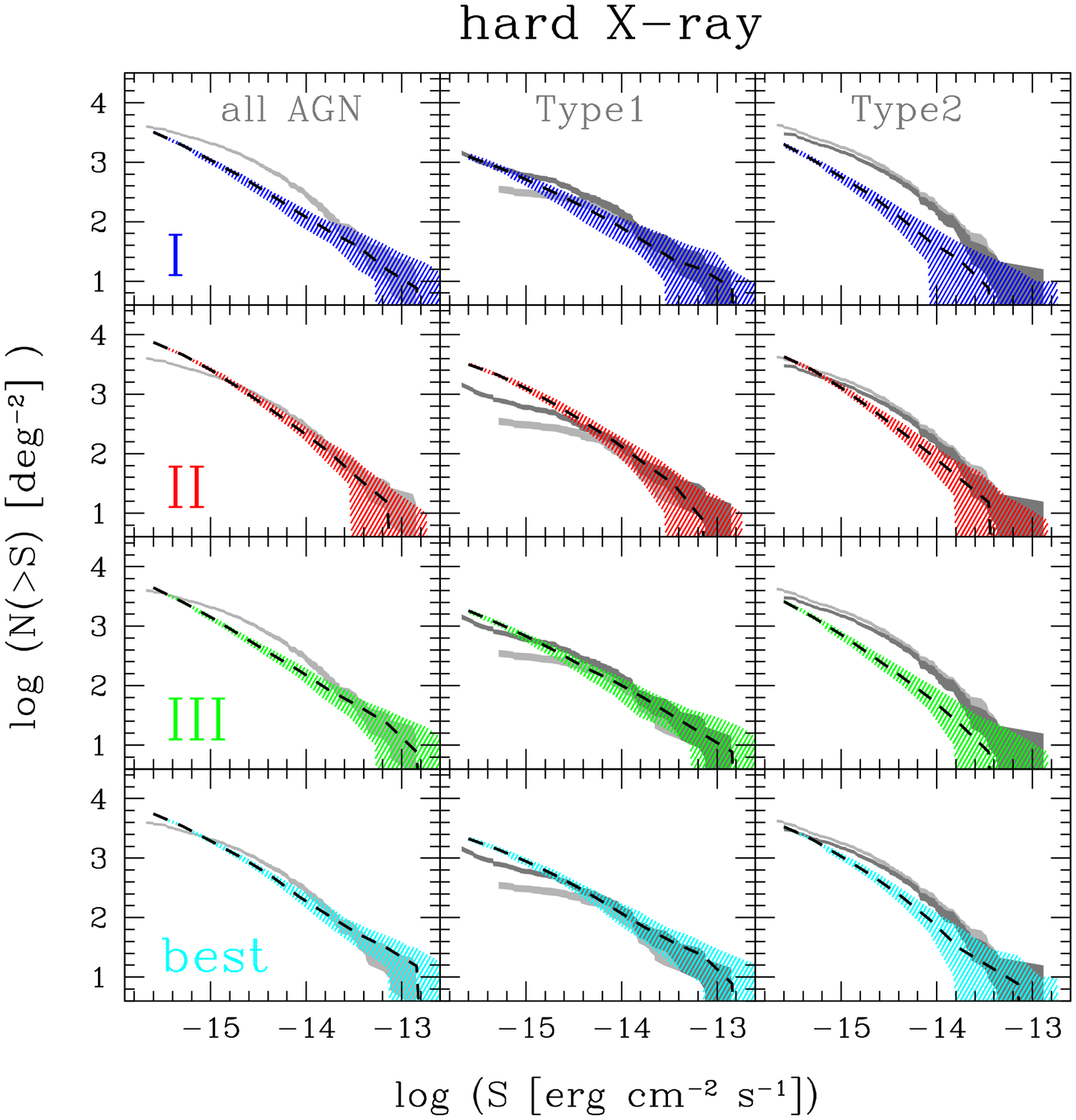}
\caption{ The predicted AGN number counts in the mock CDFs compared to
  the one determinated by \citet{bauer2004}. The left-hand and
  right-hand panels display the number counts of the AGN selected in
  the soft and hard X-ray bands, respectively. The dark and light grey
  shaded areas show the observed AGN counts obtained with two
  different classification schemes used to separate AGN from
  star-forming galaxies. Model predictions: the dashed black
    curves represents the median of all 100 CDF mocks and the bands
    indicate the 5th and 95th percentiles. Different colours
    characterize the different lightcurve models described in Section
    \ref{subsec:BH}, as indicated by the labels.}
\label{fig:logNlogS}
\end{figure*}

We have subdivided all mock AGN into type-1 and type-2, according to
their $N_{\rm H}$ absorption.  AGN with $N_{\rm
  H}<10^{22}\,\mbox{cm}^{-2}$ are classified as type-1, the more
absorbed are classified as type-2. This classification corresponds
fairly well to the optical separation into broad-line and narrow-line
AGN. All mock CDFN and CDFS pairs are extracted at large angular
separation to guarantee independent spatial correlation properties.

The left panels in Fig.~\ref{fig:cone} show the three space-like
projections of a simulated past light cone and of a mock field of view
with the same selection effects as the CDFS. The small, black dots
represent all model AGN within the cone predicted by the lightcurve
model I. The larger, green dots indicate all AGN within a mock CDFS,
placed at the centre of the box.  The panels on the right zoom in the
mock CDFS. In this case, however, the black dots show the AGN that
meet the flux-selection criteria specified above. The larger blue and
red dots show the type-1 and type-2 AGN in the mock spectroscopic
subsample defined in Section \ref{subsec:clustering}, that will be
used to compute the two-point correlation function. The size of the
red and blue dots in the upper right panel scales as the logarithm of
the AGN observed flux.


\section{Model vs. Observations} \label{sec:comparison}

In this Section, we compare the AGN number counts and spatial
clustering predicted by our model with the ones measured in the
CDFs. We quantify the dependence of our predictions on the AGN
obscuration and on the X-ray selection band. We estimate the effect of
the cosmic variance in these deep fields and investigate how robust
our conclusions are with respect to the prescription adopted for the
AGN lightcurves of individual accretion events. In order to directly
compare our predictions to observations, we use the mock AGN
catalogues constructed with the technique described in the previous
Section.

\subsection{AGN number counts}

Fig.~\ref{fig:logNlogS} shows the comparison between the AGN number
counts, $N(>S)$, predicted by our model and the ones measured in the
CDFs by \citet{bauer2004}, where $N$ is the number of AGN per unit sky
area and $S$ is their {\em observed} flux. The left-hand and
right-hand panels display the number counts of the AGN selected in the
soft and hard X-ray band, respectively. The dark and light grey shaded
areas show the observed AGN counts obtained with two different
classification schemes used to separate AGN from star-forming
galaxies, one which conservatively estimates the number of AGN and the
other which conservatively estimates the number of star-forming X-ray
sources \citep[see][for details]{bauer2004}.  The dashed black
  curves represent the median number counts computed over all 100 mock
  {\it Chandra} fields. The surrounding bands indicate the 5th and
  95th percentile. Different colours are used to characterize the
  predictions of the different lightcurve models considered in Section
  \ref{subsec:BH}.  As indicated by the labels, the model predictions
are separately compared both with the whole AGN population and with
the type-1 and type-2 ones. The width of the coloured areas is a
measure of the predicted cosmic variance. As shown in
Fig.~\ref{fig:logNlogS}, in the flux range covered by the available
observed AGN luminosity functions we recover the same results
discussed in M08. In particular, if we assume that AGN always shine at
the Eddington luminosity (model I, blue), the predicted AGN number
density is on average too low in the flux range
$\sim10^{-15}-10^{-14}$ \ergcms, especially that of the type-2
population.  Assuming a lower Eddington ratio at low redshifts, as in
our model II (red), or a decline phase of the AGN activity after an
Eddington accretion phase up to a peak mass, as in our models III
(green) and best (cyan), partly alleviates the problem.  However, at
$S\lesssim10^{-15}$ \ergcms in the soft band, i.e. in a flux range
accessible only in the X-ray selected deep fields, our model
systematically overestimates the AGN number density, irrespective of
the AGN lightcurve model, a mismatch that increases as AGN fluxes and
Eddington factor decrease.

\begin{figure}
\includegraphics[width=0.48\textwidth]{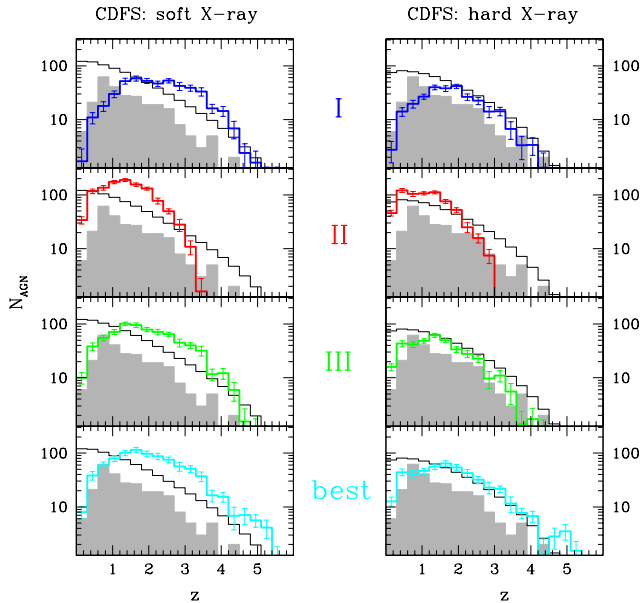}
\caption{ The redshift distribution of the AGN in our mock catalogues
  (coloured histograms). Uncertainties in the model predictions are
  computed by assuming Poisson statistics. Grey shaded histograms show
  the AGN distribution computed in the CDFS by \citet{zheng2004}. The
  solid black lines display the AGN number counts derived by
  integrating the bolometric luminosity function of
  \citet{hopkins2007}.  }
\label{fig:Zdistr}
\end{figure}

To further investigate this point, in Fig.~\ref{fig:Zdistr} we show
the redshift distribution of the AGN in our mock catalogues as a
function of the lightcurve model, as indicated by the labels. Each
model histogram has been obtained by averaging over 100 mock
catalogues. Uncertainties in the model predictions are computed by
assuming Poisson statistics. The grey shaded histograms show the
redshift distribution measured in the CDFS by \citet{zheng2004}, who
used the photometric redshifts of 342 X-ray sources, which constitute
$99\%$ of all the detected X-ray sources in the field. The solid black
lines show the AGN redshift distributions derived by integrating the
bolometric luminosity function of \citet{hopkins2007}. They can be
considered as upper limits, since this computation does not account
for the sky coverage of the fields, assuming instead a constant flux
limit for all the AGN.  As can be seen in the Figure, the faint AGN
population, overestimated by the model as shown in
Fig.~\ref{fig:logNlogS}, is distributed at all redshifts larger than
unity. The mismatch is particularly evident in the soft X-ray selected
samples.

As we did check, the number density of AGN with fluxes $\gtrsim
10^{-15}$ \ergcms predicted by all models (apart from model I) is
similar to, or slighly smaller than the observed one.  On the
contrary, all models over-predict the number density of fainter AGN
that, however, are typically excluded in the mock CDFs.  This
discrepancy can be due to one or more of the following reasons: at
$S\lesssim10^{-15}$ \ergcms, i) the mechanism triggering the BH mass
accretion is less efficient than we have assumed, ii) the accretion
time is overestimated, iii) the model fraction of obscured AGN is
underestimated.  Clearly, the model needs to be further developed
along these lines to match observations. However, for the purpose of
studying the AGN clustering in the CDFs, the over-abundance of faint
AGN in our model does not necessairly represent a problem since almost
of all of them are excluded from the spectroscopic AGN samples of the
CDFs (see below).
 
\subsection{AGN spatial clustering}\label{subsec:clustering}

We compare the spatial clustering of AGN in our mock CDFs with those
measured in the real catalogues by \citet{gilli2005} and investigate
the dependence on the AGN luminosity.  We quantify the AGN clustering
properties by means of the two-point auto--correlation function in the
real space, $\xi(r)$, using the \citet{landy1993} estimator
\begin{equation}
\xi(r)=\frac{AA(r)-2RA(r)+RR(r)}{RR(r)},
\label{eqn:1}
\end{equation}
where $AA(r)$, $RA(r)$ and $RR(r)$ are the fraction of mock AGN--AGN,
AGN--random and random--random pairs, with spatial separation, $r$, in
the range $[r-\delta r/2, r+\delta r/2]$.  The random sample is
obtained by randomly positioning objects within the same light cones
and according to the selection criteria of the AGN sample.  The
rationale behind computing $\xi(r)$ using spatial positions rather
than redshifts is that we wish to compare model predictions with the
estimates of \citet{gilli2005} and \citet{plionis2008}, in which
redshift distortions have been corrected for either by projecting the
redshift space correlation function or by inverting the measured
angular correlation function via Limber's equation.

To test whether our model is able to match the two-point correlation
functions in the CDFs measured by \citet{gilli2005}, we have extracted
mock AGN catalogues that closely mimic the spectroscopic AGN samples,
in which only objects with good optical spectra, i.e. with spectral
quality flag $Q\ge2$, are considered. For the majority of the AGN in
the CDFs, the latter condition is verified when ${\rm M_R}<25$, where
${\rm M_R}$ is the total apparent magnitude in the R band,
i.e. including the contribution of both the AGN and its host galaxy.

To extract a mock spectroscopic subsample, we have computed the R band
magnitude of all AGN in the mock {\it Chandra} Deep Fields and
rejected all objects with ${\rm M_R}>25$. In addition, since only
about half of the AGN redshifts in the {\it Chandra} Deep Fields have
been measured, we randomly diluted our sample, keeping only $50\%$ of
the mock sources. In Appendix \ref{appendix}, we describe the
procedure adopted to convert the intrinsic bolometric luminosities of
model AGN into apparent R magnitudes, given the redshift of the object
and its column density $N_H$. The observer frame R magnitudes of the
host galaxies have been obtained assuming the parametrization for dust
attenuation proposed by \citet{delucia2007}. We note that the redshift
distribution of the mock samples obtained with this procedure is
remarkably similar to those observed for the spectroscopic samples of
CDFs (e.g. Szokoly et al. 2004, Barger et al. 2003).

\begin{figure*}
\includegraphics[width=0.48\textwidth]{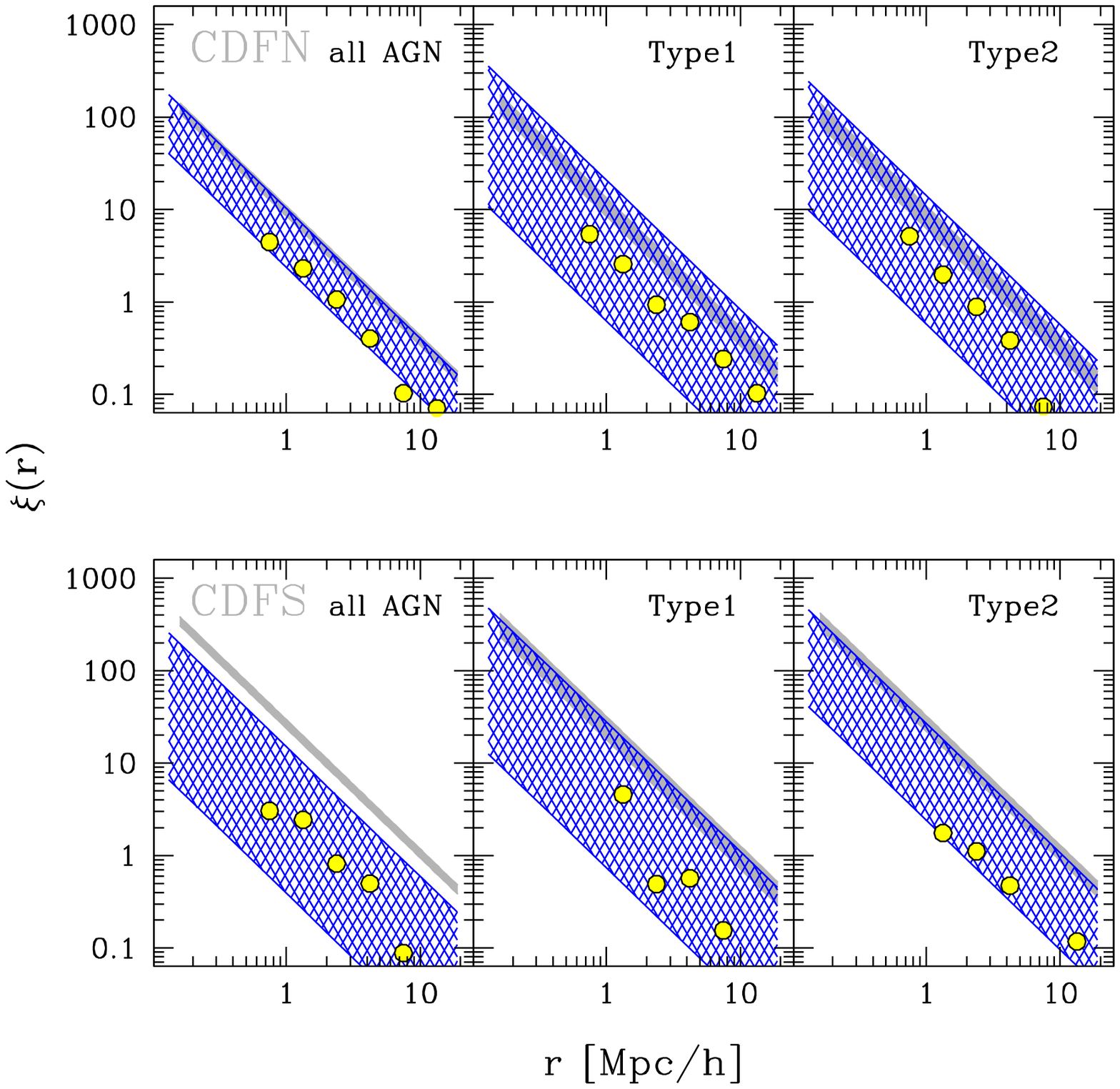}
\includegraphics[width=0.48\textwidth]{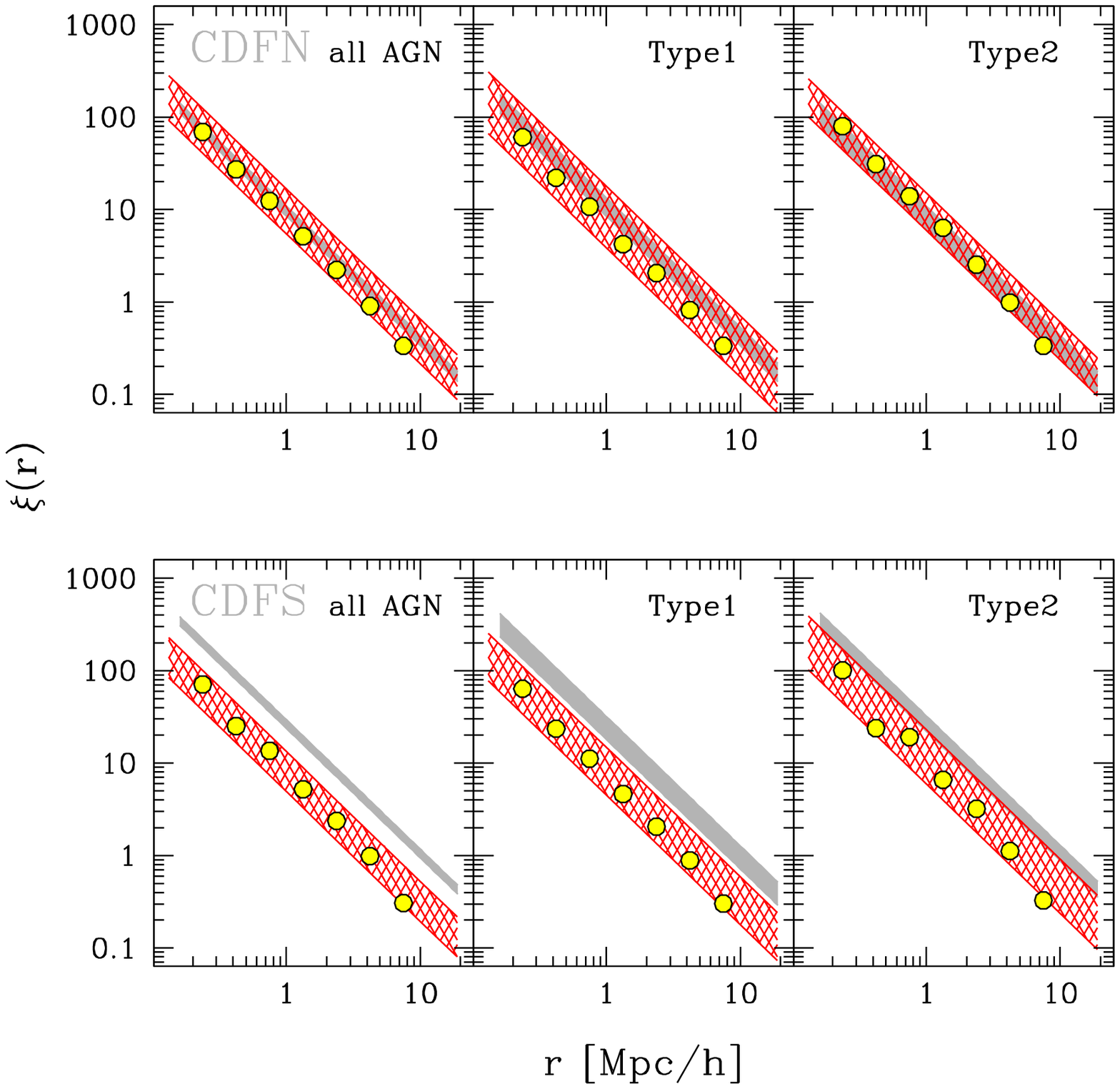}
\includegraphics[width=0.48\textwidth]{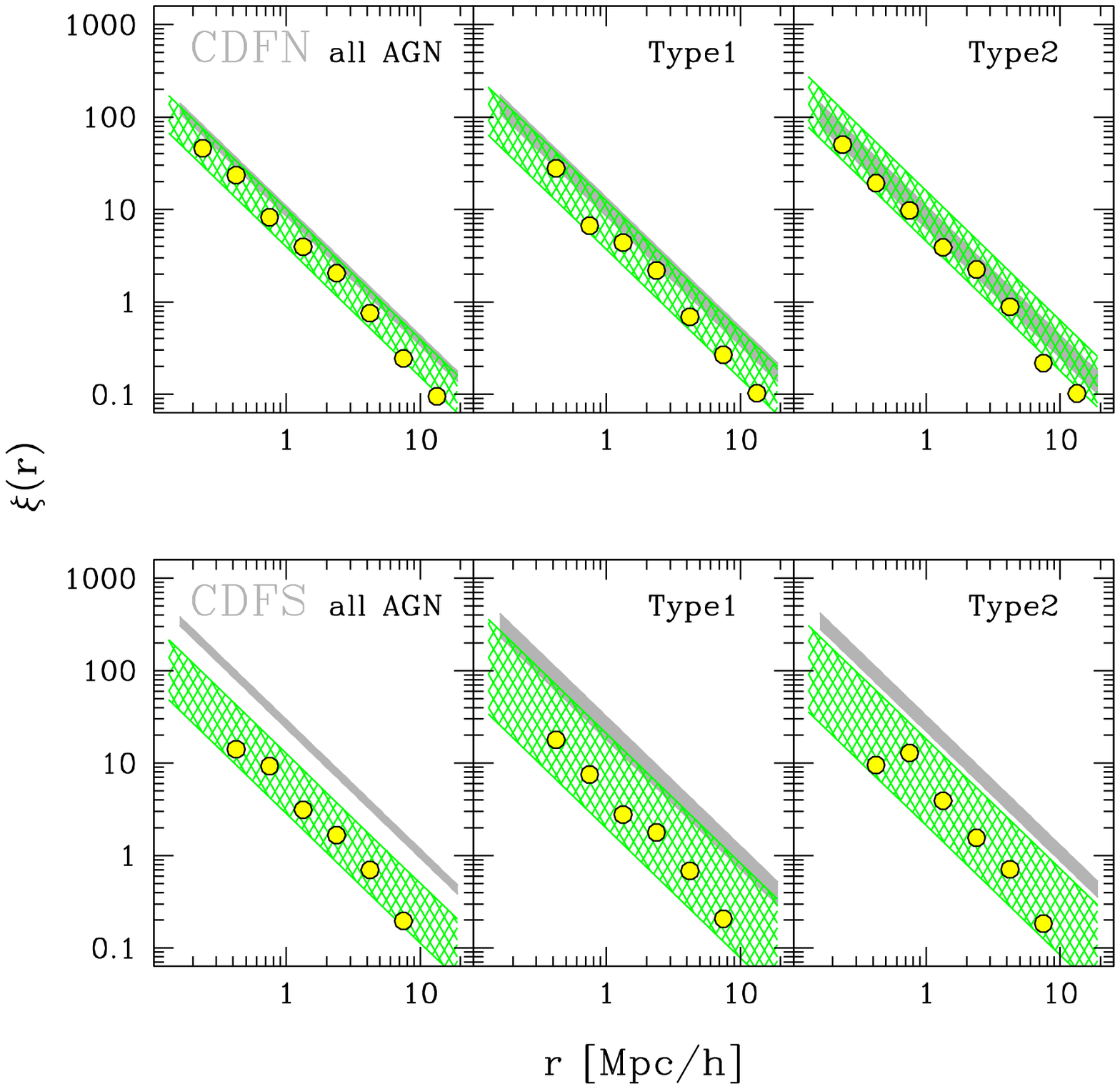}
\includegraphics[width=0.48\textwidth]{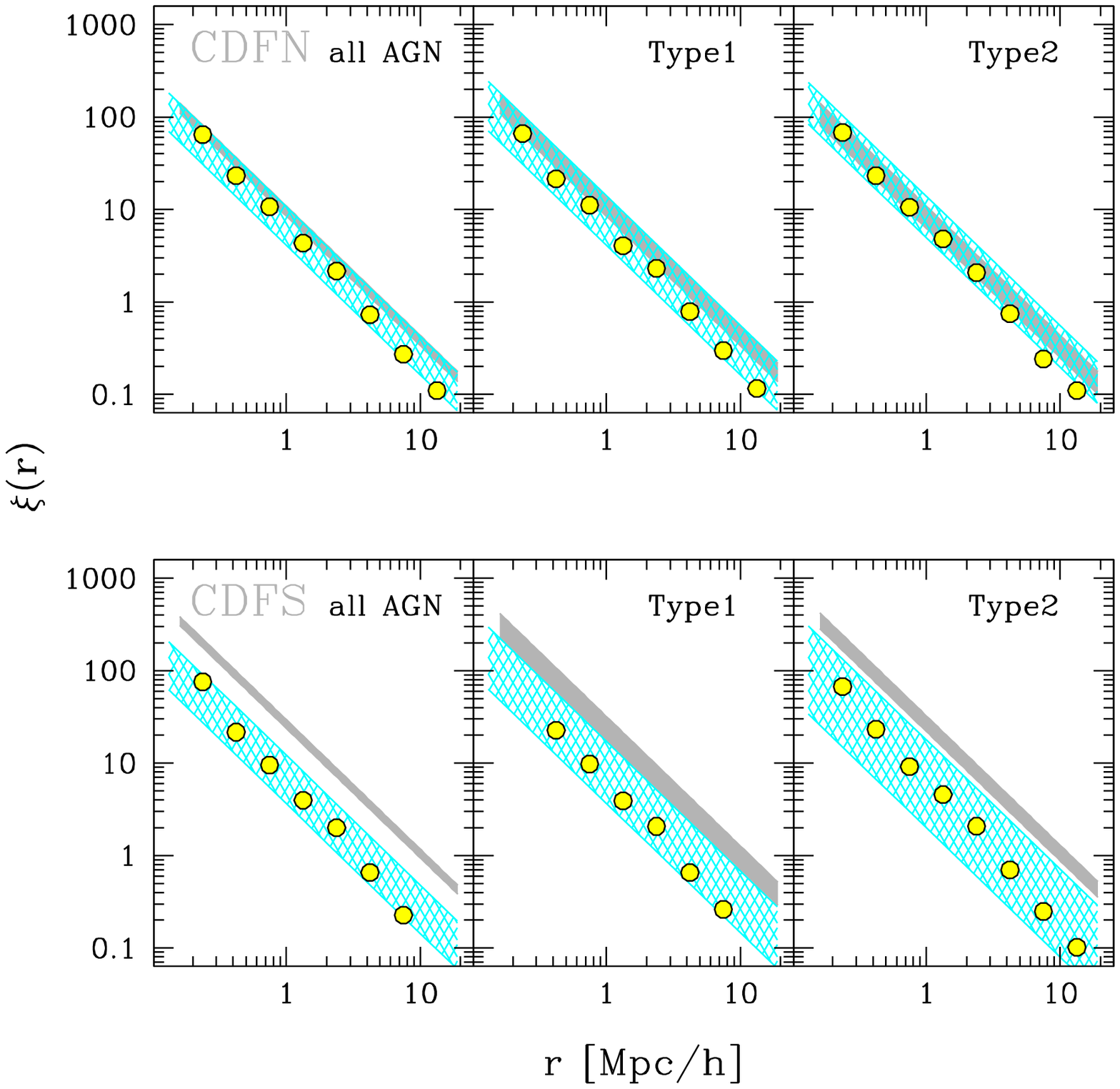}
\caption{ The four panels show the spatial two-point correlation
    function measured in 100 mock {\em Chandra} fields, as a function
  of the different lightcurve models adopted. The grey shaded areas
  have been computed using the best-fit power-law model with
    fixed slope $\gamma=1.4$ adopted by \citet{gilli2005} to the CDF
  AGN real space correlation function.  As indicated by the labels,
  the model predictions are compared both with the whole AGN
  population and with the type-1 and type-2 ones, separately. The
    yellow dots represent the correlation function of all AGN in the
    100 mock {\it Chandra} fields.  The coloured areas bracket the 5th
    and 95th percentile of the best-fit power-law to the correlation
    function in each mock sample (see Table \ref{tab}).  The bandwidth
    accounts for the different sources of uncertainties, including
    cosmic variance.  The fact that yellow dots are found within the
    shaded region indicates the adequacy of the best fit model. }
\label{fig:xi}
\end{figure*}

The grey shaded areas in the four panels of Fig.~\ref{fig:xi}
represent the power-law model two-point correlation functions that,
according to \citet{gilli2005}, best fit the correlation properties of
the AGN in the CDFs. We show the case in which the authors fixed the
slope to $\gamma=1.4$ in order to focus on the difference in the $r_0$
value between the two AGN populations, given the large errors
introduced by low number statistics.  The latter are modeled as simple
Poisson errors.  We have repeated the same best fitting procedure to
the two-point correlation function measured in each of the mock CDFs.
The result is represented by the bands of different colours. Their
width represents the field to field variance and accounts for both
sparse sampling and cosmic variance. Therefore, these errors quantify
the discrepancy between the $r_0$ in the data and the models, under
the rather strong assumption that $\gamma=1.4$.

\begin{table*}
\begin{center}
\begin{tabular}{llll}
\hline
\hline
&         & $r_0$: CDFN  &\\
\hline
& all AGN & type 1 & type 2 \\
\hline
real catalogue & $5.1\,(_{-0.5}^{+0.4})$ & $5.6\,(_{-1.0}^{+0.8})$ & $4.7\,(_{-1.0}^{+0.8})$ \\
mock I  &  $3.5 \pm 1.7 \, (0.8)$   &   $4.7 \pm 4.0 \, (1.6)$   &   $3.7 \pm 3.0 \, (1.5)$ \\
mock II  &  $5.4 \pm 2.0 \, (0.08)$   &   $5.2 \pm 2.6 \, (0.2)$   &   $5.3 \pm 1.7 \, (0.1)$ \\
mock III  &  $3.9 \pm 1.3 \, (0.2)$   &   $4.3 \pm 1.7 \, (0.5)$   &   $5.1 \pm 2.1 \, (0.4)$ \\
mock best  &  $4.1 \pm 1.3 \, (0.2)$   &   $4.7 \pm 1.9 \, (0.4)$   &   $4.8 \pm 1.7 \, (0.4)$ \\
\hline
\hline
&         & $r_0$: CDFS &\\
\hline
& all AGN & type 1 & type 2 \\
\hline
real catalogue & $10.4\,(0.8)$ & $10.1\,(_{-2.2}^{+1.8})$ & $10.7\,(_{-1.6}^{+1.3})$\\
mock I  &  $3.7 \pm 3.2 \, (1.8)$   &   $5.8 \pm 5.0 \, (2.7)$   &   $6.2 \pm 4.3 \, (2.7)$ \\
mock II  &  $4.7 \pm 1.6 \, (0.2)$   &   $4.9 \pm 2.0 \, (0.4)$   &   $6.5 \pm 2.9 \, (0.4)$ \\
mock III  &  $4.1 \pm 2.0 \, (0.6)$   &   $5.2 \pm 3.6 \, (1.2)$   &   $4.8 \pm 3.1 \, (1.5)$ \\
mock best  &  $4.2 \pm 1.7 \, (0.5)$   &   $5.1 \pm 2.6 \, (1.0)$   &   $4.7 \pm 3.1 \, (1.5)$ \\
\hline
\hline
\end{tabular}
\caption{The best fit parameters: $r_0\pm\sigma_{r_0} (\langle\rm
  err(r_0)\rangle)$, where $\xi(r)=(r/r_0)^{-1.4}$; $\sigma_{r_0}$ are
  the field-to-field variances of  $r_0$;  $\langle\rm
  err(r_0)\rangle$ are the parameters uncertainties averaged over the
  mock fields. }
\label{tab}
\end{center}
\end{table*}

The yellow dots represent the two-point correlation functions computed
using all the AGN pairs in all mock fields. The fact that they are
located within the coloured areas indicates the adequacy of the
power-law model adopted for the best fit.  As in
Fig.~\ref{fig:logNlogS}, we show our predictions for the whole AGN
population and separately for the type-1 and type-2 AGN.
 
The parameters of the best fits are listed in Table \ref{tab} together
with the errors in the form $r_0\pm\sigma_{r_0} (\langle\rm
err(r_0)\rangle)$, where $r_0$ is the best fit value, $\sigma_{r_0}$
represents the field-to-field rms and $\langle\rm err(r_0)\rangle$ is
the Poisson uncertainty on $r_0$ averaged over all mock fields.  When
comparing the errors in the mocks, that account for both sparse
sampling and cosmic variance, with the Poisson errors of
\citet{gilli2005}, we see that the error budget is dominated by cosmic
variance.  In the CDFN, the correlation length of the mock AGN is
consistent with the data. In all models the mean $r_0$ value is
smaller than the observed one. However, the difference is below
1-$\sigma$. Interestingly, our model predictions for the $r_0$ values
are in good agreement with the one estimated by considering all
extragalactic objects with measured redshifts in the CDFN, including
galaxies \citep[$r_0=4.2\pm0.4\,h^{-1}$ Mpc; see Table 2
  of][]{gilli2005}. Since galaxies make up $\sim 30$ \% of the
spectroscopic sample, this fact could be explained by assuming that
most of these galaxies actually contain a weak AGN outshone by their
host.

We did also perform a two-parameter fit as in \citet{gilli2005}. In
this case, however, the fitting procedure is not robust. Different
fitting methods provide different results and the scatter among the
best fitting values of $r_0$ and $\gamma$ is comparable, and sometimes
larger, than their formal error. The effect is larger for model I that
predicts significantly less AGN in the CDFs than the other models.
Yet, in all models explored a power-law model provides a good fit to
the measured $\xi(r)$ which, for the CDFN, is fully consistent with
the data. For example, for the model dubbed ``best'' we have obtained
$r_0=3.8 \pm 0.8$ and $\gamma=1.5 \pm 0.3 $ in the CDFN and $r_0=3.6
\pm 0.7$ and $\gamma=1.5 \pm 0.4 $ in the CDFS, where the quoted
errors represent the scatter among the mocks. These values can be
compared with the measured values $r_0=5.5 \pm 0.6$ and $\gamma=1.50
\pm 0.12$ in the CDFN and $r_0=10.3 \pm 1.7$ and $\gamma=1.33 \pm
0.14$ in the CDFS.  A two parameter fit reduces the differences
between the AGN clustering in the CDFN and CDFS. However, the lack of
robustness in the two-parameter fitting procedure and the covariance
between $r_0$ and $\gamma$ hamper a quantitative estimate. We can only
conclude that the discrepancy between the model and the observed
two-point correlation functions measured in the CDFS is smaller than
the 2-2.5 $\sigma$ difference in the correlation lengths $r_0$.

Many possible effects may help to further alleviate the tension
between model and data.  For example, we have seen that the error
budget is dominated by cosmic variance that we have estimated using
mock catalogs extracted from the Millennium Simulation. Although very
large, the computational box is still small for sufficiently rare
events. For example, it is not sufficient to contain one $z=6$ Sloan
quasar on average. And clustering statistics is more sensitive to
simulation volume than most other quantitites one typically considers.
Yet, the Millennium Simulation box can accomodate about 100
independent {\it Chandra} fields and thus the true variance should not
be significantly larger than the estimated one.  Alternatively, the
analysis of the real data might be affected by errors that have not
been accounted for in the analysis of the mock samples. For example,
the spatial two-point correlation function of \citet{gilli2005} has
been obtained from the projected one assuming a power-law model.
Possible deviations from the power-law shape would also contribute to
errors. However, according to our models, these errors should be
negligible, since the mock AGN correlation function is well
approximated by a power law.  Several examples can be worked
out. However, in order to significantly affect our results, these
hidden errors must be comparable to cosmic variance which, as we have
seen, is larger than sparse sampling error.
  
Uncertainties in model predictions provide an additional way to
reduce the discrepancy between model and data. 
For example, the clustering of
our mock AGN could be enhanced by forcing models to preferentially
populate highly biased, massive haloes.  This would increase the AGN
correlation length in both CDFN and CDFS and reduce the mismatch
between model and data. More physically motivated AGN models may
predict very different properties for AGN that populate haloes of a given mass.
This would increase the so-called stochasticity of the AGN bias and increase the 
size of the coloured regions in Fig.~\ref{fig:xi} \citep{dekel1999,sigad2000}.
However, it is not at all obvious how to achieve this task.
    
The other possibility, of course, is that the discrepancy between CDFN
and CDFS is significant and that the observed clustering of the AGN in
the CDFS is unusually large.  An indication that this may indeed be
the case is provided by the AGN two-point correlation function
recently measured in the XMM-COSMOS fields by \citet{gilli2009} which
is consistent with that of CDFN and, as we have verified, with our
model predictions, but not with that of CDFS. 

Finally, as can be seen in Fig.~\ref{fig:xi}, we stress that our
conclusions are robust with respect to the lightcurve model
assumed. Moreover, as we have verified, our results are almost
unchanged when using different assumptions in converting AGN
bolometric luminosities into optical apparent magnitudes.

\subsubsection{Luminosity dependent AGN clustering}

\citet{plionis2008} have recently investigated the clustering of the
AGN in the CDFs as a function of their luminosity.  The authors have
measured the two-point angular correlation function of the objects in
different flux-limited subsamples and then used Limber's equation to
derive the spatial clustering length $r_0$. They found a strong
dependence of $r_0$ on the median X-ray luminosity of each
flux-limited subsample in both the CDFN and CDFS and in the soft and
hard X-ray band.
 
\begin{figure}
\includegraphics[width=0.48\textwidth]{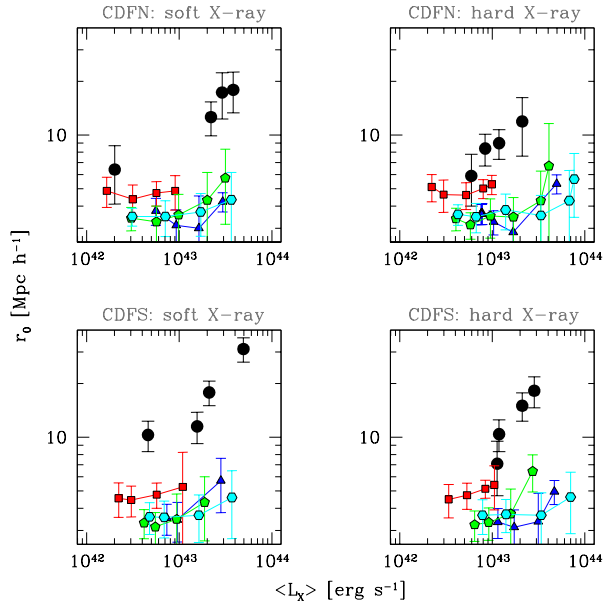}
\caption{The values of $r_0$ as a function of the median X-ray
  luminosity $\langle L_{\rm AGN, X}\rangle$ for the AGN in the mock
  CDFN (upper panels) and CDFS (bottom panels).  The results of our
  four lightcurve models are represented by different symbols (I: blue
  triangles, II: red squares, III: green pentagons, best: cyan
  hexagons) and compared with the results of \citet{plionis2008}
  (black dots).  Model predictions have been obtained by averaging
  over $100$ different mock catalogues for each lightcurve model.
    Errors show the scatter among the mock fields.}
\label{fig:xi_lmean}
\end{figure}

To investigate whether we find a similar trend in our model, we have
extracted different flux-limited subsamples from the mock {\it
  Chandra} fields, characterized by different values of $F_{\rm
  limit}$ and, therefore, by a median X-ray luminosity $\langle L_{\rm
  AGN, X}\rangle$.  The clustering length of the mock AGN in each
subsample has been estimated by fitting their spatial two-point
correlation functions with a power-law.  The results are shown in
Figure~\ref{fig:xi_lmean}, in which we plot the values of $r_0$ as a
function of $\langle L_{\rm AGN, X}\rangle$ for the AGN in the mock
CDFN (upper panels) and CDFS (bottom panels).  The results of our four
lightcurve models are represented by different symbols (model I: blue
triangles, II: red squares, III: green pentagons, best: cyan hexagons)
and compared with the results of \citet{plionis2008} (black dots).
Model predictions have been obtained by averaging over $100$ different
mock catalogues for each lightcurve model.  Errors show the
  scatter among the mock fields.

In all models the correlation length is almost constant with
luminosity, showing just a slight increase at high luminosities, in
disagreement with the strong luminosity dependence $r_0$ found by
\citet{plionis2008}.  Although small, the precise trend in the mock
catalogues depends on the lightcurve model adopted.  For instance, in
model {\em best} the dependence is quite mild, while in model II $r_0$
significantly increases already above $\langle L_{\rm AGN, X}\rangle
\sim 10^{42.5} \mbox{erg}\,\mbox{s}^{-1}$. The spread in the model
predictions makes the clustering luminosity dependence a possible
observational test to discriminate among different theoretical models
if they can be compared with larger samples in order to reduce the
size of the error bars.  The sample of AGN with measured redshift in
the 2 deg$^2$ XMM-COSMOS field represents an important step in this
direction.  Interestingly enough, the correlation length of $\sim 500$
AGN with typical X-ray luminosity of $10^{43.8}
\mbox{erg}\,\mbox{s}^{-1}$ in the 0.5-10 keV band is in the range 6-8
\Mpch (depending on whether a prominent structure at $z=0.36$ is
included or not in the sample), significantly smaller than the value
estimated by \citet{plionis2008} and in good agreement with the one
predicted by our models (Gilli et al. 2008).

We note that a global study of the clustering properties of simulated
AGN not restricted to the {\em Chandra} fields will be presented in
\citet{bonoli2008}. We anticipate here a similar result
for the luminosity dependence of AGN clustering: $r_{0}$ is found to
be only weakly dependent on luminosity, in particular in the redshift
range $z\sim 2-3$, that corresponds to the peak of the AGN number
density.

\section{Conclusions} \label{sec:conclusions}

In this paper, we modelled the AGN spatial distribution measured in
the {\em Chandra} deep fields within the framework of hierarchical
co-evolution of BHs and their host galaxies.  For this purpose, we
have applied the semi-analytic techniques developed by
\citet{croton2006}, \citet{delucia2007} and M08 to follow the
cosmological evolution of AGN inside the Millennium Simulation
\citep{springel2005}, and extracted a number of independent mock
catalogues of AGN that closely resemble the CDFS and CDFN.  Each mock
CDF catalogue has been obtained by including all AGN within a past
light cone of a generic observer that meet the same selection criteria
(field of view, flux limit, edge effects) as the real sample.  The
large volume of the Millennium Run allowed us to extract hundreds of
independent mock CDFs in which we have measured the spatial two-point
correlation function of the mock AGN in real-space.  We have ignored
redshift space distortions since these are already corrected for in
the observational estimates of \citet{gilli2005} and
\citet{plionis2008}, which we wish to compare with.

The main results of this study can be summarized as follows:

{\em (i)} The number counts of bright model AGN agree with
observations both in the soft and in the hard X-ray bands.  The
abundance of model AGN at fluxes below $\lesssim 10^{-15}$ \ergcms,
however, is larger than observed.  The amplitude of the mismatch
depends on the lightcurve model explored and on the AGN intrinsic
absorption.  In fact, our models seem to underpredict the abundance of
type-2 objects.

{\em (ii)} The number of mock AGN in the simulated CDFs in the
redshift range $1.5\lesssim z \lesssim4$ is higher than observed in
the soft X-ray band. The mismatch is less evident in the hard X-ray
band. This discrepancy in the redshift distributions is not unexpected
since, as discussed by M08, the same hybrid model considered in this
work over-predicts the abundance of faint objects with redshift in the
range $z\lesssim 4$ (see their Fig.7).

{\em (iii)} The spatial two-point correlation function predicted
  by all lightcurve models is well described by a power-law out to 20
  \Mpch.  If one set the slope $\gamma =1.4$, as in \citet{gilli2005},
  then the correlation length $r_0$ agrees, to within 1 $\sigma$, with
  that measured by \citet{gilli2005} in the CDFN once cosmic variance
  is accounted for.  On the contrary, the mock AGN in the CDFS are
  much less correlated than the real one.  In this case, the
  discrepancy in the correlation lenght is of the order of 2-2.5
  $\sigma$, depending on the lightcurve model adopted.  

{\em (iv)} The mismatch is alleviated by performing a
  two-parameter fit to the two-point correlation function.  However, a
  quantitative estimate is hampered by the lack of robustness in the
  two-parameter fitting procedure which results from low number
  statistics.  The tension between model and data is further
  alleviated by possible observational errors that are not properly
  accounted for and by model uncertainties. Overall, one expects that
  the discrepancy between the observed and modeled $\xi(r)$ is smaller
  than the 2-2.5 $\sigma$ mismatch in the correlation lengths quoted
  previously.

{\em (v)} The agreement between correlation functions in the
  XMM-COSMOS field \citep{gilli2009} and in the CDFN which, as we have
  shown, is well reproduced by our AGN models suggests that the AGN
  clustering in the CDFS is indeed unusually high.

{\em (vi)} The models predict that the clustering amplitude depends
little on the luminosity of AGN, in disagreement with the strong
dependence found by \citet{plionis2008} but in agreement with the
measurements of the clustering of luminous AGN in the recently
complied XMM-COSMOS catalogue \citep{gilli2009}.

Precise predictions for the luminosity dependence of the AGN
clustering depend on the adopted theoretical models, and their present
mutual agreement merely reflects the still large field-to-field
variance. Therefore, one can hope that measuring the AGN clustering
properties as a function of their luminosity in larger datasets could
help discriminating among the models. Furthermore, going beyond the
spatial AGN autocorrelation function, the analysis of the
cross-correlation between AGN and galaxies in the next generation
all-sky surveys at $z\ge 1$, like EUCLID or ADEPT, will place strong
constraints on modern semi-analytic models, thereby shedding light on
the complicated mechanisms that regulate the co-evolution of AGN and
galaxies.


\section*{acknowledgments}
We thank M. Roncarelli, N. Sacchi and A. Bongiorno for very useful
discussions and M. Plionis for making available his results. FM thanks
the Max-Planck-Institut f\"ur Astrophysik for the kind hospitality. We
acknowledge financial contribution from contracts ASI-INAF I/023/05/0,
ASI-INAF I/088/06/0 and ASI-INAF I/016/07/0. SB acknowledges the PhD
fellowship of the International Max Planck Research School in
Astrophysics, and the support received from a Marie Curie Host
Fellowship for Early Stage Research Training. We would also like to
thank the anonymous referee for helping to improve and clarify the
paper.

\bibliographystyle{mn2e} \bibliography{bib}

\appendix
\section{From AGN intrinsic luminosities to
  observed magnitudes} \label{appendix}

To convert the intrinsic bolometric luminosities of the model AGN,
$L_{\rm bol}$, into absorbed apparent R-band magnitudes, given the AGN
redshift and $N_{\rm H}$, we make the following steps. First, we use
the bolometric correction given by \citet{hopkins2006} to get the AGN
intrinsic B-band luminosity, $L_{\rm B}$. Then, we get the
monochromatic unabsorbed R-band luminosity, assuming:
\begin{equation}
L^{\rm UNABS}_\nu=L_{B,\nu}\left(\frac{\nu}{\nu_B}\right)^{-0.44}\, ,
\end{equation}
where
\begin{equation*}
L_{B,\nu}=\frac{\lambda^2}{c}L_{B,\lambda}\sim\frac{\lambda^2}{c}\frac{L_B}{\Delta\lambda_B}\,
, 
\end{equation*}
$\nu_B=c/(445 \rm nm)$, $\nu=(1+z)\nu_R=(1+z)c/(658 \rm nm)$,
$\Delta\lambda_B\sim100 \rm nm$ and $c$ is the speed of light.  The
absorbed monochromatic luminosity can be obtained with the following
equation:
\begin{equation}
L^{\rm ABS}_\nu=L^{\rm UNABS}_\nu\times10^{-0.4\,A}\, ,
\end{equation}
where 
\begin{multline}
  A=A_V\left(1+\frac{1}{3.1}(0.000843x^5-0.02496x^4+\right.\\\left.0.2919x^3-1.815x^2+6.83x-7.92)\right)\, ,
\end{multline}
$x=\lambda^{-1}$ in $\mu m^{-1}$ and $A_V=5\times10^{-22}N_{\rm H}$ \citep{gaskell2007}.

Finally, to get the apparent R magnitude in the observer frame, we
use:
\begin{equation}
R_{\rm AB}=8.9-2.5\log(f_\nu/Jy)\, ,
\end{equation}
where $f_\nu$, the monochromatic flux expressed in units of Jansky, is:
\begin{equation}
f_\nu=(1+z)\frac{L^{\rm ABS}_\nu}{4\pi d_L(z)^2}\, ,
\end{equation}
and $d_L(z)$ is the luminosity distance.

To get the total R-band magnitudes of mock objects, the AGN magnitude
computed as described above is finally combined with that of the host
galaxies obtained by \citet{delucia2007}.

\label{lastpage}

\end{document}